\pdfoutput=1
\documentclass[
    10pt,floatfix,a4paper,
    prl,
    amsfonts,amssymb,amsmath,
    showpacs,
    twocolumn,
    superscriptaddress,
    groupedaddress
]{revtex4}
\usepackage{graphicx}
\usepackage{amssymb,amsmath}
\usepackage{amsmath}
\usepackage{fullpage, epic, eepic, psfrag}

\newcommand{\Figref}[1]{Fig.~\ref{#1}}

\newcommand{\ca}{{\cal A}}
\newcommand{\cb}{{\cal B}}

\newcommand{\geh}{{\gamma^{\textrm{eh}}}}

\begin{document}

\title{Laser-like vibrational instability in rectifying molecular conductors}
\author{Jing-Tao \surname{L\"u}}
\email{jtlu@nanotech.dtu.dk}
\affiliation{DTU Nanotech, Dept. of Micro and Nanotechnology, Technical University of Denmark, {\O}rsteds Plads, Build. 345E, DK-2800 Kongens Lyngby, Denmark}
\author{Per \surname{Hedeg{\aa}rd}}
\affiliation{Niels Bohr Institute, Nano-Science Center, University of Copenhagen, Universitetsparken 5, 2100 Copenhagen {\O}, Denmark}
\author{Mads \surname{Brandbyge}}
\affiliation{DTU Nanotech, Dept. of Micro and Nanotechnology, Technical University of Denmark, {\O}rsteds Plads, Build. 345E, DK-2800 Kongens Lyngby, Denmark}
\date{\today}

\begin{abstract}
We study the damping of molecular vibrations due to electron-hole pair excitations in
donor-acceptor(D-A) type molecular rectifiers. At finite voltage additional non-equilibrium electron-hole pair excitations involving both electrodes become possible, and contribute to the stimulated emission and absorption of phonons.  We point out a generic mechanism for D-A molecules, where the stimulated emission can dominate beyond a certain voltage due to inverted position of the D and A quantum resonances. This leads to current-driven amplification (negative damping) of the phonons similar to laser-action.
We investigate the effect in realistic molecular rectifier structures using first principles calculations.
\end{abstract}
	

\pacs{73.63.-b, 85.65.+h, 72.10.Di, 73.40.Gk}
\maketitle

In a seminal paper from 1974, Aviram and Ratner proposed an electronic rectifier based on a single organic molecule\cite{ArMa.1974}. Akin to the $p$-$n$ junction in solid-state electronics their design involved a donor and an acceptor group bridged by a tunnel barrier(D-A). Ever since, the interest in molecule-based electronic operations, such as rectification\cite{Metzger1999,ElOcKo.05,DiHiLe.09}, has increased\cite{Ta.2006}. Recently, focus has been on the interaction between the electronic current and the atomic dynamics in molecular conductors\cite{GaRaNi.2008,Hihath2010,NoMaHiYoTa.2010}. It is well known that a fraction of the electronic current exchange energy with the molecular vibrations leading to Joule heating\cite{GaRaNi.2007,HuChDa.2007,IoShOp.2008,WaCoTo.10}.
However, instabilities such as conductance switching or break-down occur in experiments\cite{SmUnRu.2004,TsTeKu.2006,ScFrGa.2008,DiHiLe.09,MoLi.10} in the high voltage regime, which still poses open
questions to theory. Recent theoretical work has indicated that current-induced forces can constitute an important channel of energy exchange between electrons and ions {\em different} from Joule heating, and lead to destabilization or runaway behavior of molecular contacts\cite{DuMcTo.2009,LuBrHe.2010}.

In this paper we point out a generic mechanism leading to instabilities in D-A molecular rectifiers. We show that the electron-hole pair excitation by the atomic vibrations, known as electronic friction\cite{PEPE.80a,HETU.1995}, may not necessarily damp out the vibrational energy in biased D-A molecular rectifiers. Instead, we can get negative friction or phonon amplification beyond a certain voltage. This happens when the stimulated emission of phonon dominates over absorption processes due to a population inversion, similar to what happens in a laser.

We first outline the theoretical basis, then illustrate the effect by a simple two-level model, before presenting our first-principles calculations on realistic systems relevant for experiments.

\textit{Theory.---} Within the harmonic approximation, ignoring mode-mode coupling,
we can calculate the current-induced excitation of a given vibrational mode(phonon) using the rate equation approach for the phonon population, $N$,
\begin{equation}
 \dot{N}={\cb}(N+1)-{\ca}N\,.
	\label{eq:powerab}
\end{equation}
Here all phonon emission(absorption) processes are described by the rate, ${\cb}$(${\ca}$), and we have
the steady-state occupation,
\begin{equation}
N_{ss}=\frac{1}{(\ca /\cb)-1}\,.
\end{equation}
In equilibrium $N_{ss}=n_B(\hbar\Omega)$, where $n_B$ is the Bose occupation at the phonon energy($\hbar\Omega$), corresponding to $\ca=e^{\hbar\Omega/k_B T}\cb > \cb$. However, out of equilibrium, when an electronic current is present, we may obtain an occupation $N_{ss}\gg n_B$. Most importantly, we may obtain a 'laser'-type instability, namely if $\cb$ approaches $\ca$.
Using Fermi's golden rule, we can calculate the rates resulting from the coupling to the electronic subsystem in the presence of current,
\begin{equation}
{\cb}={\displaystyle\frac{2\pi}{\hbar}\sum_{i,f}} |\langle f|M|i\rangle|^2 F_i (1-F_f)\delta(\Delta\varepsilon_{fi}+\hbar\Omega)\,.
\end{equation}
Here $M$ is the electron-phonon coupling between initial($i$) and final($f$) electronic states with occupations $F$,
$\Delta\varepsilon_{fi}=\varepsilon_f - \varepsilon_i$. $\ca$ is given by the corresponding expression with
$\Omega\rightarrow -\Omega$.

For an applied voltage, $V$, we employ the usual non-equilibrium setup with left($L$) and right($R$) electrodes at
different chemical potential ($\mu_L-\mu_R=eV$), corresponding to Fermi distributions $n_F^\alpha$, $\alpha\in\{L,R\}$.
We consider low temperatures, where $k_B T\ll eV, \hbar\Omega$. In this limit $\cb$ is only non-zero if $eV > \hbar\Omega$, and it can be written as,
\begin{equation}
{\cb}=\int_{\mu_R+\hbar\Omega}^{\mu_L}\!\Lambda_{LR}(\varepsilon,\varepsilon-\hbar \Omega)d\varepsilon,
\label{eq:B}
\end{equation}
introducing the electrode-resolved, coupling-weighted DOS for the electron-hole pairs,
\begin{equation}
\Lambda_{\alpha\beta}(\varepsilon,\varepsilon')=\frac{1}{2\pi\hbar}\textrm{Tr}\left[MA_\alpha(\varepsilon) MA_\beta(\varepsilon') \right]\,,
\label{eq:Lambda}
\end{equation}
given by the spectral densities, $A_\alpha$, of scattering states (DOS) originating in electrode $\alpha$.
In the same approximation the corresponding expression for ${\ca}$ becomes,
\begin{eqnarray}
\ca &=& \int_{\mu_R-\hbar\Omega}^{\mu_L}\! \Lambda_{LR}(\varepsilon,\varepsilon+\hbar\Omega)d\varepsilon \nonumber \\
 &&+ \int_{\mu_R-\hbar\Omega}^{\mu_R}\! \Lambda_{RR}(\varepsilon,\varepsilon+\hbar\Omega)d\varepsilon \nonumber \\
 &&+ \int_{\mu_L-\hbar\Omega}^{\mu_L}\! \Lambda_{LL}(\varepsilon,\varepsilon+\hbar\Omega)
 d\varepsilon.
 \label{eq:A}
\end{eqnarray}

We shall now assume that we are dealing with a molecule with a donor level with energy $\varepsilon_D$ which is most strongly coupled to the left lead, while an acceptor level with energy $\varepsilon_A$ is primarily coupled to the right lead. In this case $\Lambda_{LR}(\varepsilon,\varepsilon')$ will have a resonance as a function of $\varepsilon$ close to $\varepsilon_D$ and a resonance as a function of $\varepsilon'$ around $\varepsilon_A$. The instability occurs when $\cb>\ca$. Inspecting \eqref{eq:B}, we see how $\cb$ becomes large if $\varepsilon_D$ is in the energy window $[\mu_R+\hbar\Omega;\mu_L]$, {\em and} $\varepsilon_D-\hbar\Omega \approx \varepsilon_A$. In this case $\ca$ will be far from its optimum values since the integration intervals in \eqref{eq:A} are far from the resonances of  $\Lambda_{LR},\Lambda_{RR},\Lambda_{LL}$.
The instability can also be described as the electron-hole pair damping rate\cite{PEPE.80a,FrPaBr.2007},
\begin{equation}
\geh=\ca-\cb\,
\label{eq:geh}
\end{equation}
going negative. In principle other damping mechanisms such are coupling to bulk phonons, could be added to $\geh$.

\begin{figure}[htpb]
\begin{center}
	\includegraphics[scale=0.4]{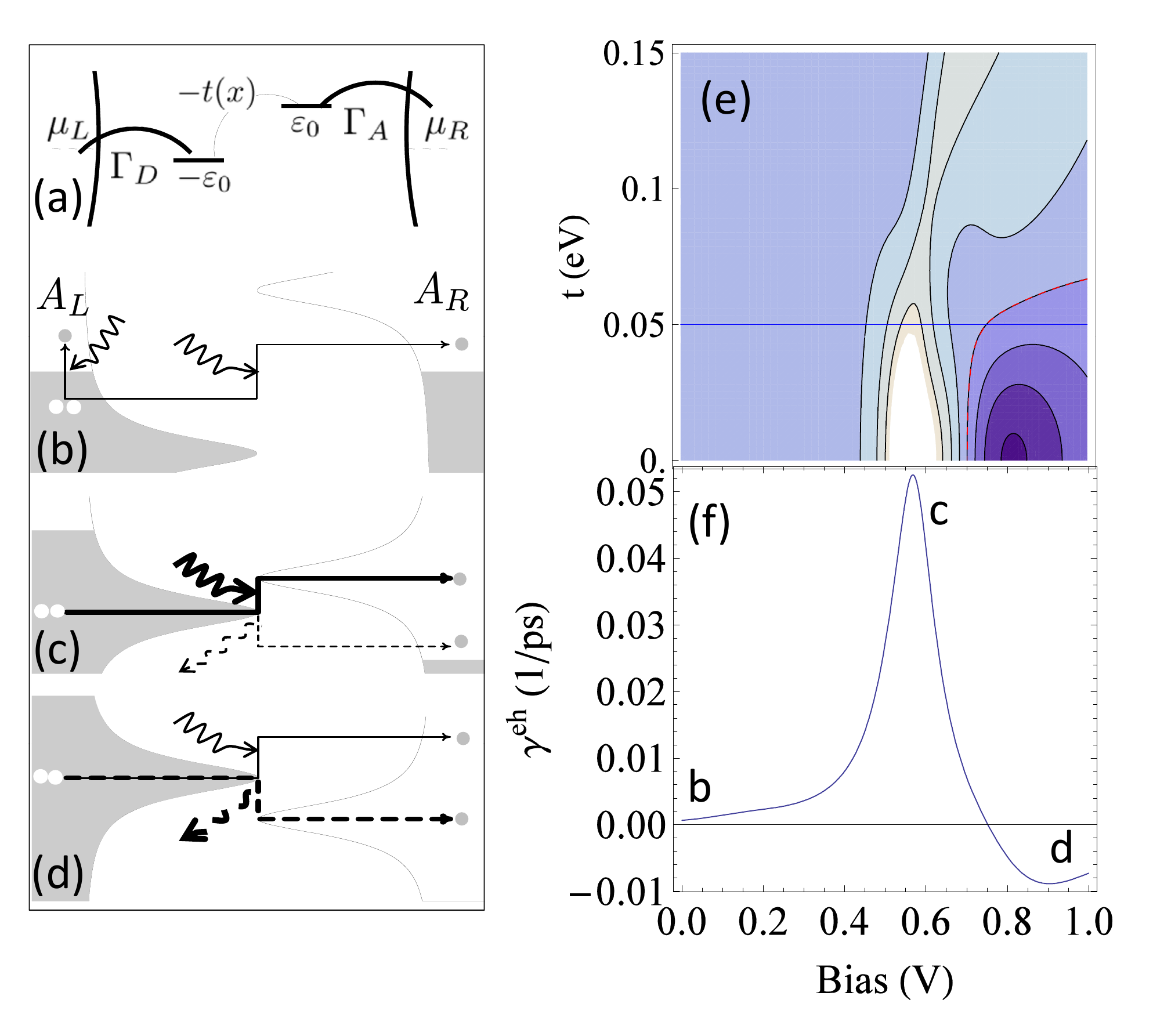}
\end{center}
\caption{(Color online) (a) Simple two-level model for the donor-acceptor (DA)
system. (b-d)D- and A-DOS and filling(grey) at $T=0$. The D(A)-levels follow the
chemical potential, $\mu_L$($\mu_R$) of the closest L(R) electrode.
(b) For zero
voltage $eV\equiv\mu_L-\mu_R=0$ only absorption processes around
$\mu_L=\mu_R\equiv\mu_0$ are possible.
Similar excitations from the right electrode are not shown.
(c) Absorption dominates for a voltage where $\varepsilon_A-\varepsilon_D\approx \hbar\Omega$ and $\geh$ is
maximal.
(d) Population inversion for larger voltage when,
$\varepsilon_D-\varepsilon_A\approx \hbar\Omega$: Emission dominates and
$\geh<0$.
(e) Contours of $\geh(V)$ for varying $t$ (dark is negative), $\varepsilon_0=0.35$eV, $\Gamma_1=0.3$eV, $\Gamma_2=0.1$eV,
$m=0.005$eV, $\hbar\Omega=20$meV, $\mu_0=-0.2$eV.
(f) $\geh(V)$ for $t=0.05$eV
showing the generic behavior (maximum followed by negative values). }
\label{fig:contour}
\end{figure}
\textit{Simple two-level model.---}  In Fig.~\ref{fig:contour}a, we consider the simplest model with
a donor-level($\varepsilon_D=-\varepsilon_0$) which couple with the left electrode($\Gamma_D$), and an
acceptor($\varepsilon_A=\varepsilon_0$) coupling to right elec\-trode($\Gamma_A$).
If the DA hopping matrix element, $t$, is small, these levels will follow the nearby
elec\-trode chemical potential\cite{StTaBr.3}. We introduce an electron-phonon coupling corresponding
to modulation of the DA hopping 'distance', $m\propto dt/dx$.

The processes involved at $T=0$, are shown in \Figref{fig:contour}b-d. In
equilibrium($\mu_L=\mu_R=\mu_0$) electron-hole pair generation both {\em intra}- and
{\em inter}-electrode lead to damping(absorption), see \Figref{fig:contour}b. The
high D(A)-DOS around $\varepsilon_D$($\varepsilon_A$) will dominate processes
when both DOS peaks are within the 'voltage window', $[\mu_R;\mu_L]$,
and have a different filling. As the bias is increased the
$\varepsilon_D$($\varepsilon_A$) moves up(down). A maximum damping is expected
when the filled D-states are separated from the empty A-states by
$\varepsilon_A-\varepsilon_D\approx\hbar\Omega$ making absorption ($\ca$) a
dominating process, see Fig.~\ref{fig:contour}c. More interestingly, as the
bias is further increased so $\varepsilon_D-\varepsilon_A \approx\hbar\Omega$
we have a situation of 'population inversion': Electronic transitions from $\varepsilon_D$(filled) to $\varepsilon_A$(empty) now makes phonon emission dominate over absorption, see Fig.~\ref{fig:contour}d. This
corresponds to the $\Lambda_{LR}(\varepsilon,\varepsilon-\hbar\Omega)$-term and thus $\cb$ to be dominating.
For fixed $m$ and sufficiently small $t$ we reach a situation where the friction becomes negative when $\varepsilon_A-\varepsilon_D\approx\hbar\Omega$, \Figref{fig:contour}e. For big $D-A$ coupling the effect disappears as the coupling makes bonding/anti-bonding DA-states more relevant. For a particular choice of $t$ the generic behavior of a maximum followed by negative values of $\geh$ is clearly seen, \Figref{fig:contour}f.

\begin{figure}[htpb]
\begin{center}
	\includegraphics[scale=0.4]{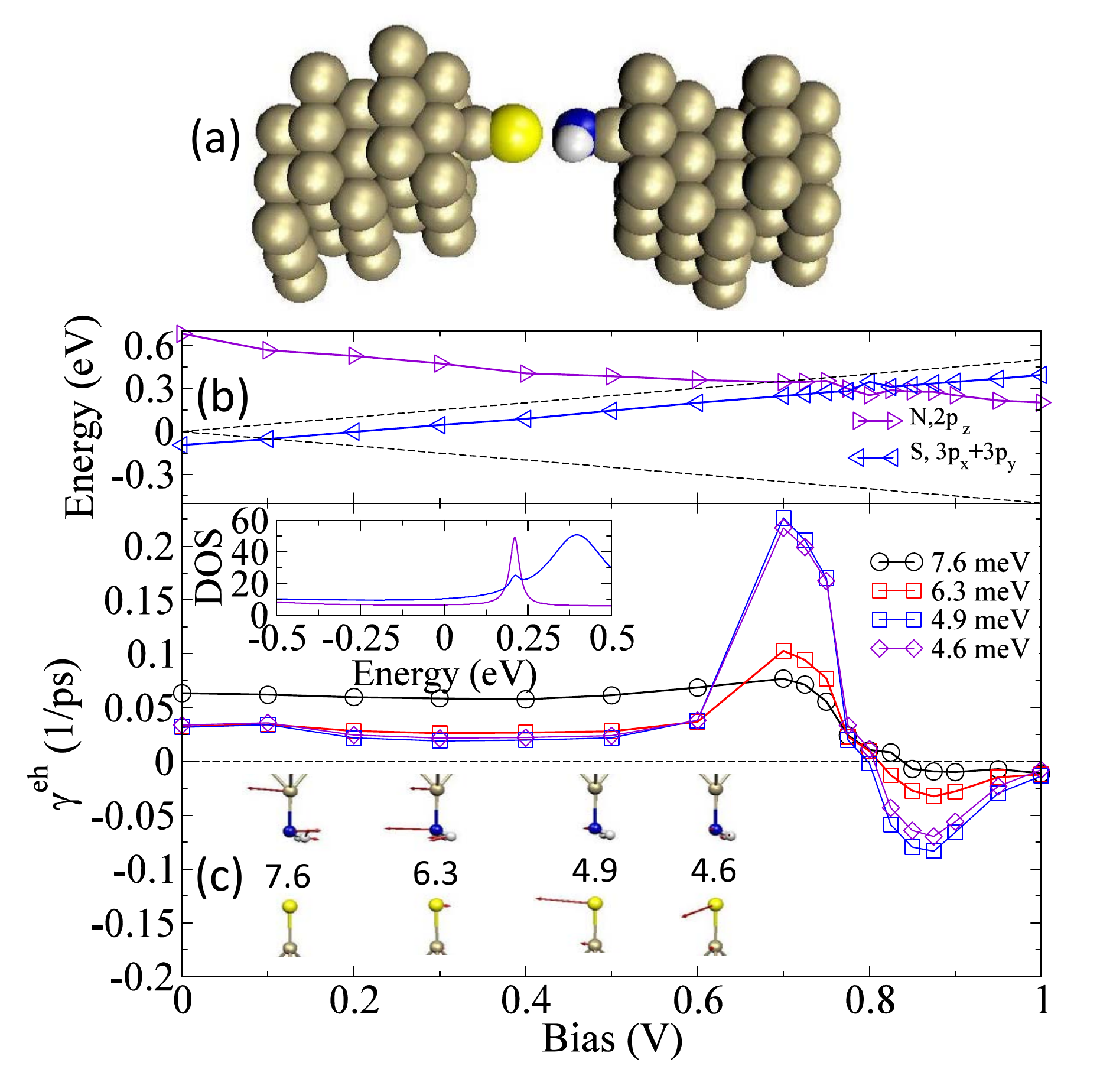}
\end{center}
\caption{(Color online)(a) Molecular STM junction: S atom adsorbed on the left adatom("tip"), the right("sample") adatom has adsorbed an NH$_2$ molecule.
(b) Positions of the D ($3p_x,3p_y$ of S) and A levels
($2p_z$ of N) as a function of bias. The dashed lines indicate $\mu_L,\mu_R$.
Inset: Left/right DOS for $1$V bias.
(c) Electron-hole damping($\geh$) as a function of voltage bias for unstable phonon modes(inset) showing
$\geh<0$ at high bias. }
\label{fig:tip}
\end{figure}

We conclude that the phonon amplification instability, $\geh < 0$, is similar to a two-level atomic laser.
In the presence of bias, the D state, filled by
electrons, is located at an energy above the empty A state.
In this case a phonon gets amplified by the stimulated emission accompanying electronic transitions
from $\varepsilon_D$ to $\varepsilon_A$. This is analogous to the population inversion and stimulated emission of photons in lasing.

\begin{figure}[htpb]
\begin{center}
	\includegraphics[scale=0.4]{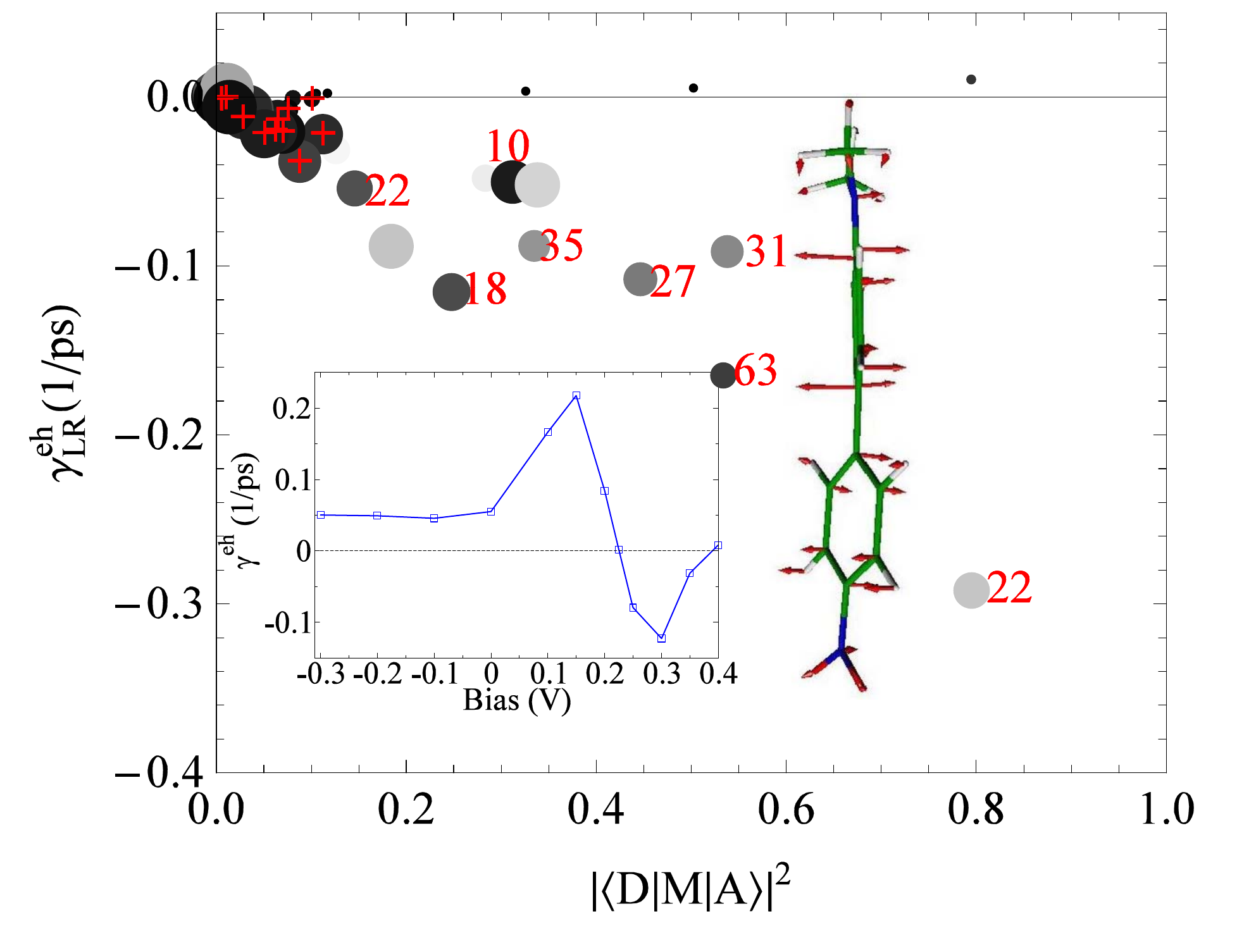}
\end{center}
\caption{(Color online) Correlation between the {\em inter}-electrode damping rate,
$\gamma^{\rm{eh}}_{\rm{LR}}$ (incl. only $LR$ terms in $\ca$, $\cb$), and
phonon-coupling matrix element between the D and A-states.
Each dot represents one phonon mode. Larger dot has smaller mode frequency, and
darker one has smaller $\gamma^{\rm{eh}}_{\rm{LL}}+\gamma^{\rm{eh}}_{\rm{RR}}$. Modes displaying
$\geh<0$ for $V<0.4$V are marked with cross line or frequency. Insets: The $63$meV mode, and bias dependence of its damping rate.}
\label{fig:phase}
\end{figure}

\textit{First-principles calculations.---} We have performed NEGF-DFT based transport calculations
\cite{Soler.02,computationaldetails} on two types of experiments, where the
instability appear. In the first example, we note that the tip in an STM can be functionalized by adatoms\cite{RuBeSt.93} or molecules\cite{NiItUm.05,ScFrAr.11} enabling control of both voltage and hopping ($\sim t$). This offers
an ideal experimental test of the instability effect. In Fig.~\ref{fig:tip}a we consider a junction formed by two gold adatoms with S and NH$_2$ adsorbed, representing sample and functionalized tip. The $3p_{x,y}$ orbitals of S, located just below the $\mu_L$, serves as D and follows $\mu_L$. The $2p_z$ orbital of N follows $\mu_R$, and serves as A, see Fig.~\ref{fig:tip}(b). In \Figref{fig:tip}(c) we show the
$\geh(V)$ for the unstable modes with negative damping(insert). The behavior is similar to the two-level model, and
involve transverse adsorbate motion as selected by the symmetry of the D and A.
\begin{figure}[htpb]
\begin{center}
	\includegraphics[scale=0.5]{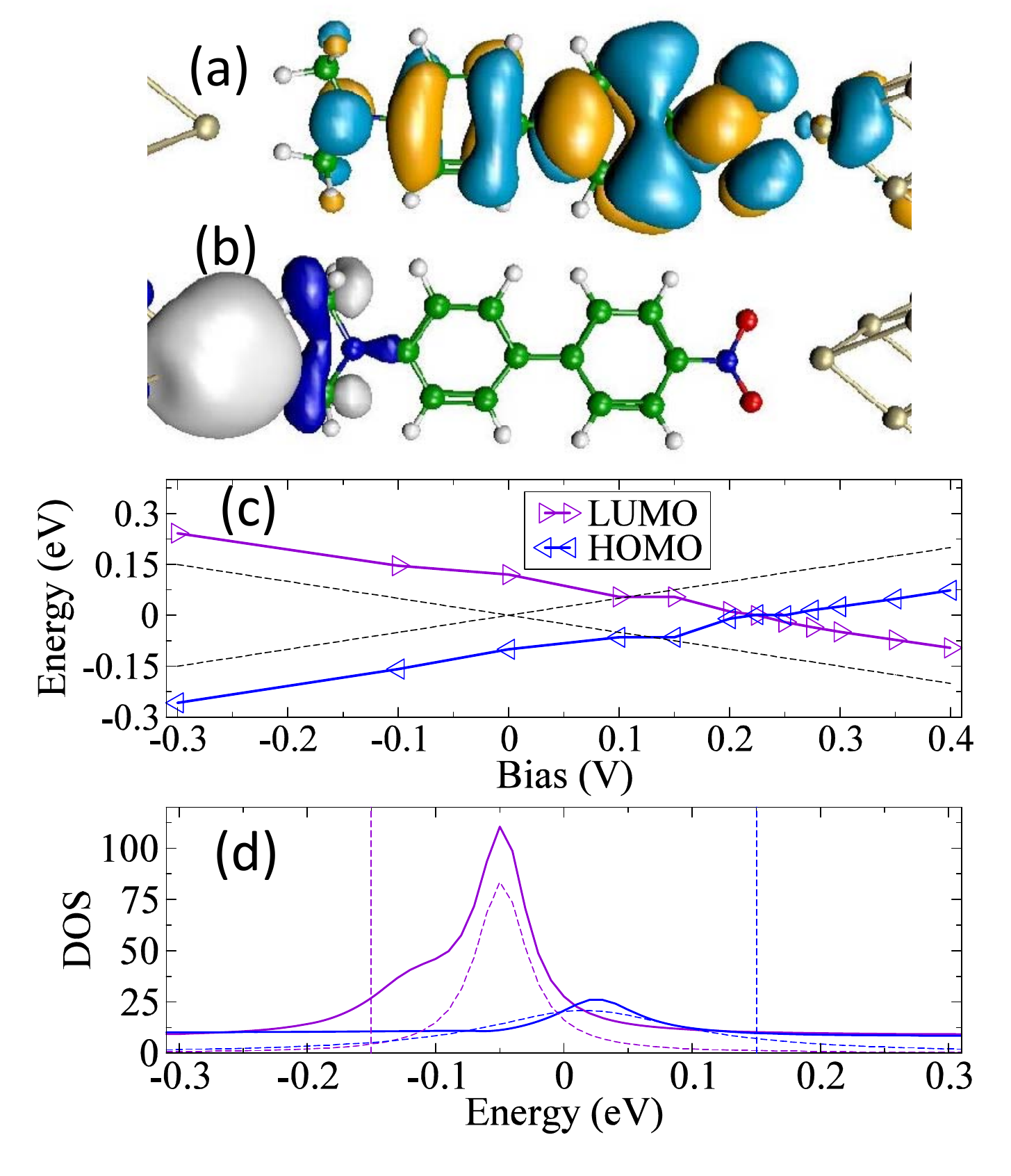}
\end{center}
\caption{(Color online)(a)-(b) The A(LUMO) and D(HOMO) states at $0.3$V ($\mu_L=0.15$V,$\mu_R=-0.15$V shown as vertical lines).
(c) The position of D and A levels(DOS peaks) as a function of bias. (d) The $A_L$, $A_R$ at $0.3$V.
The dashed lines include only contribution from the single D or A
orbitals, while the solid lines include all contributions.}
\label{fig:level}
\end{figure}
In the next example the D-A behavior is tied to the surface anchoring
groups of the 4-dimethylamino-4'-nitrobiphenyl molecule bridging Au(111) electrodes\cite{FoHoMc.8,ZoKiCuPaHuScEr.2010}.
We identify the D, A states from the molecular projected Hamiltonian\cite{StTaBr.3}.
The A-state(Fig.~\ref{fig:level}a) penetrating to the left has a $\pi$-character, while the D-state with $s$-character(Fig.~\ref{fig:level}b) exists near the
left electrode. Again $A_L$($A_R$) mainly involves $D$($A$)(\Figref{fig:level}c) tied to $\mu_L(\mu_R)$(\Figref{fig:level}d), and the unstable mode displays the typical bias dependent $\geh$. The
D,A state symmetries implies that the mode involves transverse vibrations, see inset of
Fig.~\ref{fig:phase}. In \Figref{fig:phase} we show the correlation
between the {\em inter}-electrode damping terms ($\Lambda_{LR}$), and the electron-phonon matrix
element between the D,A states, $\langle D|M|A \rangle$.  It is seen that a larger matrix
element leads to higher negative {\em inter}-electrode damping terms, while the competition with the
{\em intra}-electrode damping (and possibly other damping mechanisms) eventually determines if the mode become unstable.
We note that the $63$meV mode becoming unstable at $\sim 0.2$V has a frequency out side the electrode phonon bands. Thus we do not expect additional harmonic damping due to these.

\textit{Conclusions.---} We have discussed how stimulated amplification of phonons
becomes possible in molecular rectifiers once the donor level is lifted higher than the acceptor
by the applied bias. This is akin to population inversion leading to 'lasing', and may cause
instabilities such as switching or contact
disruption for a certain voltage\cite{DiHiLe.09}. On the other hand, current-induced cooling may also be possible\cite{IoShOp.2008}.
Both effects draw some analog with
the current-induced negative damping and cooling of a nanomechanical
oscillator\cite{BlImAr.2005}.

We thank Prof. Jauho for comments, the Lundbeck foundation for financial support(R49-A5454),
the Danish Center for Scientific Computing(DCSC) and Dir. Henriksens Fond
for providing computer resources.



\begin{thebibliography}{30}
\bibitem[{com()}]{computationaldetails}
\bibinfo{note}{We employ the SIESTA,TranSIESTA/Inelastica methods [23,21],
  similar parameters as in Ref. 21, and the $\Gamma$-point approximation .}
\end{thebibliography}

\end{document}